\documentclass[twocolumn,showpacs]{revtex4-1}
\usepackage{amsfonts,amssymb,amsmath}
\usepackage{array,dcolumn}
\usepackage{graphicx}

\begin{document}

\title{Persistent chimera states in nonlocally coupled phase oscillators}

\author{Yusuke Suda}
\author{Koji Okuda}
\affiliation{Division of Physics, Hokkaido University, Sapporo 060-0810, Japan}

\date{\today}

\begin{abstract}
Chimera states in the systems of nonlocally coupled phase oscillators
 are considered stable in the continuous limit of spatially distributed
 oscillators.
However, it is reported that in the numerical simulations without taking
 such limit, chimera states are chaotic transient and finally collapse
 into the completely synchronous solution.
In this Rapid Communication, we numerically study chimera states by
 using the coupling function different from the previous studies and
 obtain the result that chimera states can be stable even without taking
 the continuous limit, which we call the persistent chimera state.
\end{abstract}


\maketitle

The behavior of coupled oscillator systems can describe various pattern
formations in a wide range of scientific fields~\cite{b:Kuramoto,
b:PRK}.
In the systems of nonlocally coupled identical oscillators, there often
appears a strange phenomenon called the chimera state, which is
characterized by the coexistence of coherent and incoherent domains,
where the former domain consists of phase-locked oscillators and the
latter domain consists of drifting oscillators with spatially changing
frequencies~\cite{KB2002, AS2004, SK2004, AS2006, AMSW2008, OA2008,
OA2009, MLS2010, OWM2010, WOYM2011, WO2011, OMHS2011, ORHMS2012,
TNS2012, HMRHOS2012, OOHS2013, MTFH2013, SSKG2014, ZZY2014, HSK2015,
SSA2008, MVSLM2014, XKK2014, RRHSG2014}.
This interesting phenomenon was first discovered in the system of
nonlocally coupled phase oscillators obeying the evolution equation
\begin{equation}
 \frac{\partial}{\partial t} \theta(x,t) = \omega - \int dx' G(x-x') \,
  \Gamma( \theta(x,t) - \theta(x',t) )
  \label{eq: continuous phase oscillators}
\end{equation}
with $2\pi$-periodic phases $\theta(x)$ on a finite interval $x\in[0,1]$
under the periodic boundary condition, a smooth $2\pi$-periodic coupling
function $\Gamma$, and the kernel $G(y)=(\kappa/2)\exp(-\kappa|y|)$,
where a constant $1/\kappa$ denotes the coupling range~\cite{KB2002}.
Recently, similar spatiotemporal patterns have been found in various
systems using, e.g., the logistic maps~\cite{OMHS2011, ORHMS2012},
R\"{o}ssler systems~\cite{ORHMS2012}, and FitzHugh-Nagumo
oscillators~\cite{OOHS2013}.

In the study of the chimera state, the system, Eq.~(\ref{eq: continuous
phase oscillators}), with the sine coupling~\cite{SK1986}
\begin{equation}
 \Gamma (\phi) = -\sin( \phi + \alpha )
  \label{eq: sine coupling}
\end{equation}
is particularly important because of its simplicity and generality.
In fact, this coupling function was used also in the first discovery of
the chimera state~\cite{KB2002}.
For numerical simulations, we usually discretize Eq.~(\ref{eq:
continuous phase oscillators}) into such a form as Eq.~(\ref{eq: phase
oscillators}).
In the simulations of such discretized systems, we can confirm that
chimera states are surely stable in the continuous limit
$N\rightarrow\infty$.
However, the stability of chimera states in finitely discretized systems
is questioned.
In fact, it is reported that when $N$ is finite, chimera states with the
sine coupling are chaotic transient and finally collapse into the
completely synchronous solution~\cite{WOYM2011, WO2011, RRHSG2014}.

Recently, Ashwin and Burylko proposed the weak chimera similar to the
chimera state, which is defined by the coexistence of
frequency-synchronous and -asynchronous oscillators in the systems of
coupled indistinguishable phase oscillators but is not necessarily
spatially structured as coherent and incoherent domains~\cite{AB2015}.
They studied the weak chimera in some types of networks composed of the
minimal number of oscillators with the Hansel-Mato-Meunier coupling,
Eq.~(\ref{eq: HMM coupling}), and demonstrated that the weak chimera can
be persistent (non transient).
In this Rapid Communication, we study chimera states in the systems of
nonlocally coupled phase oscillators with the Hansel-Mato-Meunier
coupling by numerical simulation, and demonstrate that it is possible
for persistent chimera states to appear.

As a model, we consider a ring of $N$ identical nonlocally coupled phase
oscillators described as
\begin{equation}
 {\dot \theta}_j(t) = \omega + \frac{1}{2R} \sum^{j+R}_{k=j-R}
  \Gamma( \theta_j(t) - \theta_k(t) )
  \label{eq: phase oscillators}
\end{equation}
with $2\pi$-periodic phases $\theta_j$ ($j=1,\ldots,N$).
This model corresponds to a spatially discretized version of
Eq.~(\ref{eq: continuous phase oscillators}) with a constant kernel
within a certain range.
The natural frequency $\omega$ of the oscillators can be set to zero
without loss of generality, and the nonlocal coupling range $R$ needs to
satisfy $1<R<(N-1)/2$.
In this Rapid Communication, we fix $R/N\sim0.35$.
As the coupling function $\Gamma(\phi)$, we choose the
Hansel-Mato-Meunier coupling~\cite{HMM1993}
\begin{equation}
 \Gamma(\phi) = -\sin( \phi + \alpha ) + r \sin(2\phi),
  \label{eq: HMM coupling}
\end{equation}
where $\alpha$ is the phase lag parameter of the fundamental harmonic
component and $r$ is the amplitude ratio of the second harmonic
component.
For $r=0$, Eq.~(\ref{eq: HMM coupling}) recovers the sine coupling,
Eq.~(\ref{eq: sine coupling}).
In the systems of globally coupled phase oscillators, it is known that
such higher harmonic components in the coupling function are responsible
for a rich variety of synchronous patterns excluded by the sine
coupling~\cite{HMM1993, Daido1992, Okuda1993, Daido1996, KK2001,
ABM2008}.
Therefore we expect that also in the systems of nonlocally coupled phase
oscillators with Eq.~(\ref{eq: HMM coupling}), we could observe new
chimera patterns excluded by the sine coupling.

\begin{figure}[tb]
 \centering
 \includegraphics[width=86truemm]{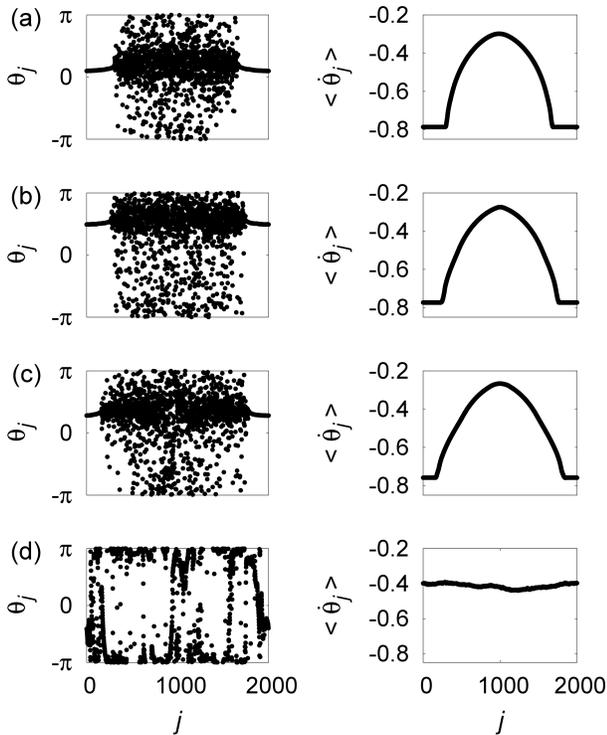}
 \caption{\label{fig: chimera states}
 Results of numerical simulation of Eq.~(\ref{eq: phase oscillators})
 with $N=2000$ and $\alpha=1.46$.
 In each row, the left panel shows the snapshot of phase $\theta_j$, and
 the right panel shows the profile of the average frequency
 $\langle\dot{\theta}_j\rangle$ with $T=5000$ and $t_{\text{rel}}=2000$
 in Eq.~(\ref{eq: average frequency}).
 For (a)~$r=0.001$, (b)~$r=0.03$, and (c)~$r=0.06$, chimera states are
 observed, while they are not observed for (d)~$r=0.12$.
 }
\end{figure}
First, we consider the case of sufficiently large $N$ corresponding to
the continuous limit.
Figure~\ref{fig: chimera states} shows the results of numerical
simulation of Eq.~(\ref{eq: phase oscillators}) with Eq.~(\ref{eq: HMM
coupling}) for several $r\geq0$.
For all the simulations of the present Rapid Communication, we used the
fourth-order Runge-Kutta method with time interval $\Delta t=0.01$.
In Fig.~\ref{fig: chimera states}, we fix $\alpha=1.46$, for which
chimera states are observed in the case of the sine coupling ($r=0$).
As initial conditions, we used
\begin{equation}
 \theta_j(0) =
  6\exp\left[ -30 \left( \frac{j}{N} - \frac{1}{2} \right)^2 \, \right]
  R_j,
  \label{eq: initial condition}
\end{equation}
where $R_j\in[-1/2,1/2]$ is a uniform random number, which is so close
to a chimera state as to assist its emergence~\cite{AS2006}.

In our simulation, chimera states are observed for $r<0.073$ as shown in
Figs.~\ref{fig: chimera states}(a)-\ref{fig: chimera states}(c).
The phase pattern (left panels) is clearly separated into coherent and
incoherent domains, which is characteristic of the chimera state.
From the right panels of this figure, we can see that the average
frequency
\begin{equation}
 \langle \dot{\theta}_j \rangle(T) = \frac{1}{T}
  \int^{t_{\text{rel}}+T}_{t_{\text{rel}}} \dot{\theta}_j(t) \, dt
  \label{eq: average frequency}
\end{equation}
of each oscillator in the coherent domain is almost constant, where $T$
is the measurement time and $t_{\text{rel}}$ is the relaxation time,
while the frequency in the incoherent domain varies continuously.
For $r\geq0.073$, chimera states gradually disappear as  $r$ increases.
In addition, for $r\geq0.110$, chimera states are not observed, but each
oscillator evolves almost independently, where the average frequency
seems to converge to a constant value in the limit of
$T\rightarrow\infty$, though the frequency in Fig.~\ref{fig: chimera
states}(d) still exhibits some fluctuations due to a finite $T$.
The survey of these behaviors is depicted in Fig.~\ref{fig: phase
diagram}.
\begin{figure}[tb]
 \centering
 \includegraphics[width=86truemm]{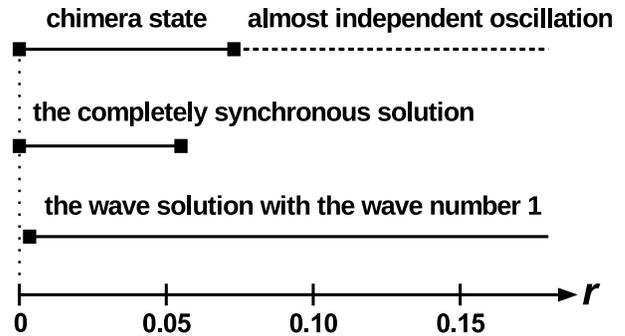}
 \caption{\label{fig: phase diagram}
 Phase diagram of stable solutions to Eq.~(\ref{eq: phase oscillators})
 with Eq.~(\ref{eq: HMM coupling}) at $\alpha=1.46$  in the continuous
 limit ($N=2000$).
 Horizontal lines denote the stability regions of each solution.
 We have not clearly determined the transition between the chimera state
 and almost independent oscillation yet.
 Note that the stability region of the wave solution with the wave
 number $k=1$ does not cover $r=0$.
 }
\end{figure}

From the linear stability analysis, it is found that the completely
synchronous solution $\theta_1=\theta_2=\cdots=\theta_N$ to
Eq.~(\ref{eq: phase oscillators}) with Eq.~(\ref{eq: HMM coupling}) is
stable for $r<(\cos\alpha)/2$ ($\simeq0.055$ at $\alpha=1.46$).
Moreover, chimera states also appear to be stable in this parameter
region [see Figs.~\ref{fig: chimera states}(a) and \ref{fig: chimera
states}(b)].
However, it is reported that when $N$ is finite, chimera states at $r=0$
are transient and finally collapse into the completely synchronous
state~\cite{WO2011}.
We below confirm whether these chimera states, particularly for $r>0$,
are transient or really stable even when $N$ is finite.

\begin{figure}[tb]
 \centering
 \includegraphics[width=80truemm]{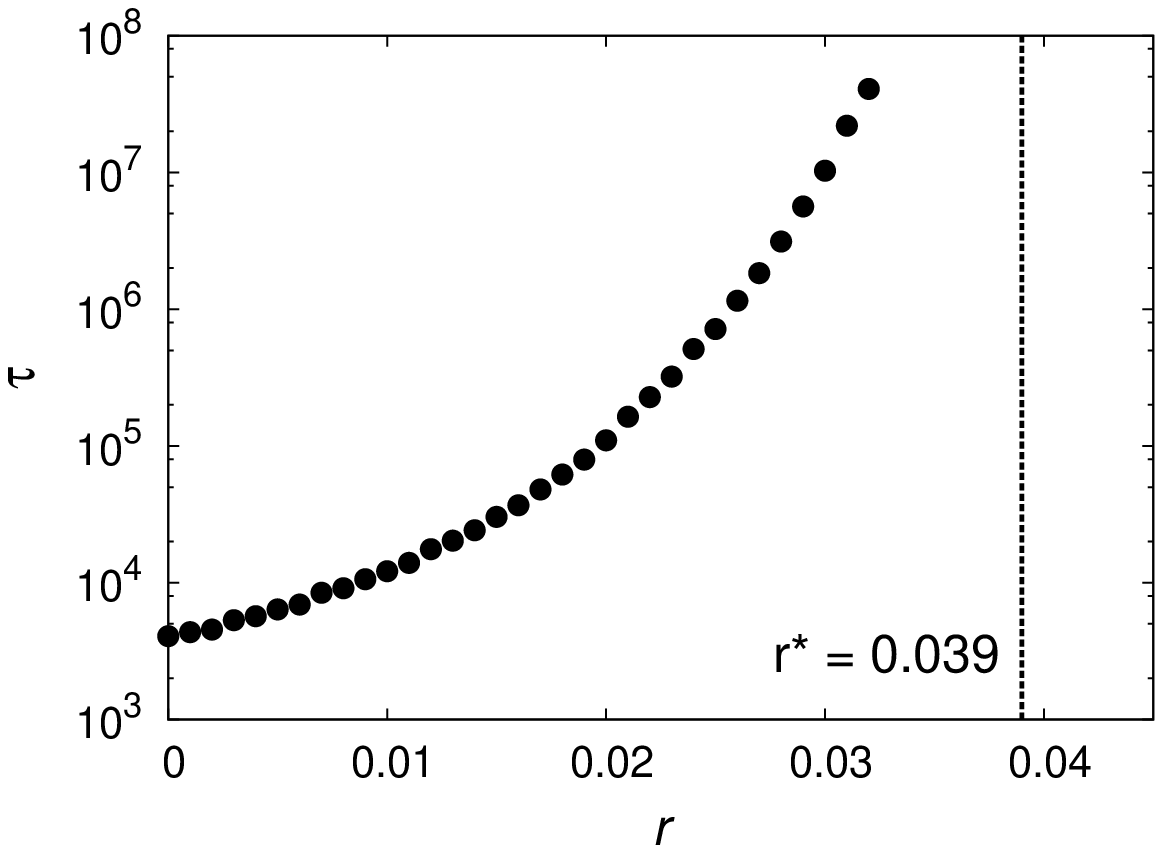}
 \caption{\label{fig: average lifetime}
 Average lifetime $\tau$ of the chimera state as varying $r$ at the
 parameters $N=30$ and $\alpha=1.46$.
 A point in the figure is the average over $1000$ simulations from
 different initial conditions obeying Eq.~(\ref{eq: initial condition}).
 }
 \vspace{5truemm}
 \centering
 \includegraphics[width=86truemm]{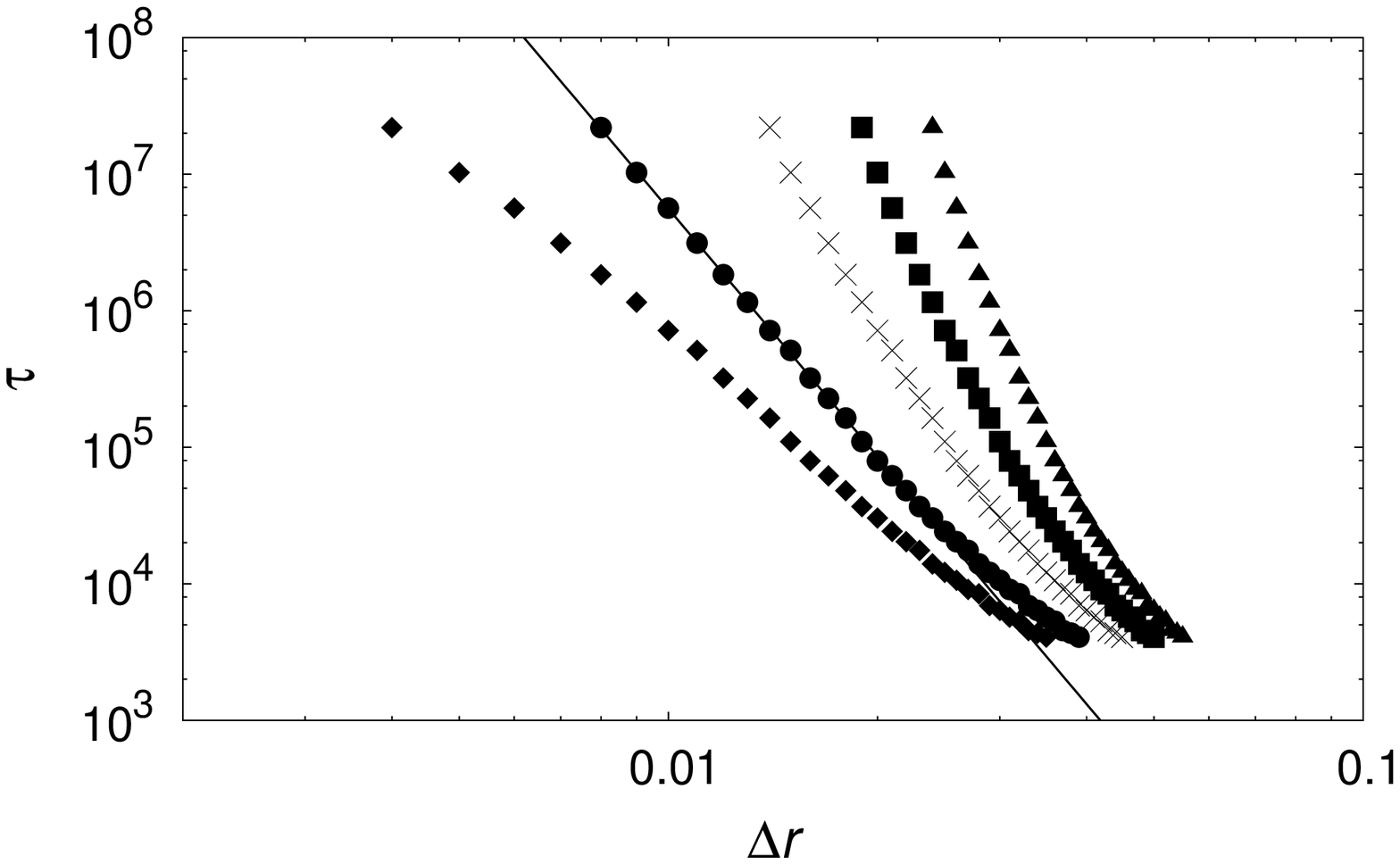}
 \caption{\label{fig: critical r}
 Log-log plot of the data $(\Delta r,\tau)$ for $r^{\ast}=0.035$
 (diamond), $0.039$ (circle), $0.045$ (cross), $0.050$ (square) and
 $0.055$ (triangle).
 The data for $r^{\ast}=0.039$ are fitted linearly by the least squares
 method (black line), where we used only the data $\tau\ge300000$ to
 obtain better linearity for this fitting.
 }
\end{figure}
Figure~\ref{fig: average lifetime} shows the average lifetime $\tau$ of
the chimera state for $N=30$, as increasing $r$ from $0$ to
$(\cos\alpha)/2\simeq0.055$.
Here we regard the lifetime of the chimera state as the time at which
the completely synchronous state appears, i.e., the global order
parameter
\begin{equation}
 Z(t) = \left| \frac{1}{N} \sum_{k = 1}^{N} e^{i\,\theta_k} \right|
  \label{eq: global order parameter}
\end{equation}
reaches $Z(t)=1$.
As for the chimera state in the finite $N$ cases, it should be noted
that it is difficult to judge the emergence of the chimera state,
because the spatial position of the chimera state does not stay still
but fluctuates~\cite{OWM2010}, in particular, more violently as $N$
becomes smaller.
In fact, in the case of $N=30$, we could not observe the characteristic
profile of the average frequency as in the right panels of
Figs.~\ref{fig: chimera states}(a)-\ref{fig: chimera states}(c).
However, we observed that the coherent domain exists in the phase
snapshots as in the left panels of Figs.~\ref{fig: chimera
states}(a)-\ref{fig: chimera states}(c), which convinces us of the
emergence of the chimera state.

For Fig.~\ref{fig: average lifetime}, it should be noted that there is a
possibility that chimera states collapse into a stable solution other
than the completely synchronous state.
From the linear stability analysis, we can show that the wave solution
$\theta_i=\theta_1+2\pi k(i-1)/N$~\cite{SSA2008, XKK2014} with the wave
number $k=1$ is also stable for $r\geq0.003$ (Fig.~\ref{fig: phase
diagram}).
This implies that chimera states may collapse into the wave solution.
However, we never observed such collapse in our simulations from $1000$
different initial conditions, Eq.~(\ref{eq: initial condition}), at each
$r$.

As $r$ is increased, the average lifetime $\tau$ increases
monotonically, and appears to diverge to infinity at a certain
$r=r^{\ast}$.
Assuming some values as $r^{\ast}$, we obtain Fig.~\ref{fig: critical r}
by the log-log plot of the data $(\Delta r,\tau)$, where $\Delta r\equiv
r^{\ast}-r$.
From this figure, we can assume the power law
\begin{equation}
 \tau \propto (\Delta r)^{-\zeta}
  \label{eq: critical r}
\end{equation}
and we determine $r^{\ast}\simeq0.039$ from the best linear fitting of
the data.
Since $r^{\ast}<(\cos\alpha)/2\simeq0.055$, this implies that there
exists a parameter region where the chimera state (with infinite
lifetime) and the completely synchronous state are bistable even in the
finite $N$ cases.
However, we cannot exclude the possibility of $r^{\ast}=(\cos\alpha)/2$,
because it is difficult to obtain the exact value of $r^{\ast}$ due to
divergent simulation time.

Next, we investigate the chimera state of $N=30$ for $r>(\cos\alpha)/2$,
where the completely synchronous state is unstable.
The possibility that chimera states appear in the region without the
stable completely synchronous state differs from the case of the sine
coupling.
In this region, chimera states cannot collapse into the completely
synchronous state.
Though the wave solution with $k=1$ is stable in this region, we never
observed that chimera states collapse into the wave solution within our
maximum simulation time $t=2\times10^8$.
Therefore, the collapse of the chimera state should not occur if other
stable non-chimera solutions do not exist.
Though we searched for stable non-chimera solutions other than the wave
solution by extensive numerical simulations, we could not find any such
solutions.
From the above results, we conclude that, in a certain range of $r>(\cos
\alpha)/2$, the chimera state and the wave solution are bistable, and
the chimera state can be persistent (non-transient) even in the finite
$N$ cases.
This result is consistent with $\tau\rightarrow\infty$ for $r>r^{\ast}$,
as seen in Fig.~\ref{fig: average lifetime}.

\begin{figure}[tb]
 \centering
 \includegraphics[width=86truemm]{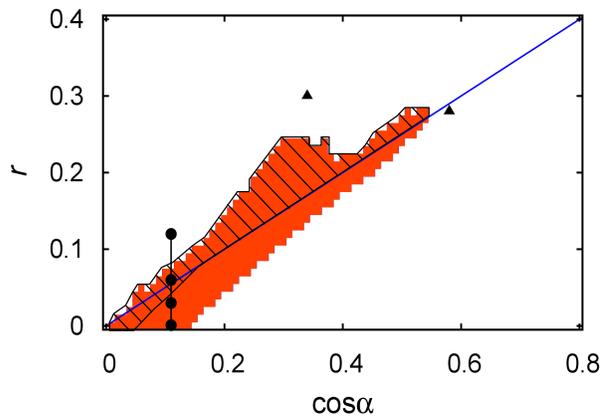}
 \caption{\label{fig: stability region}
 (Color online)
 Stability region of the chimera state (red) on the $(\cos\alpha,r)$
 plane in the continuous limit ($N=2000$).
 The hatched region corresponds to the persistent chimera state, that
 is, the stability region of the chimera state in the case of $N=30$.
 The blue line denotes $r=(\cos\alpha)/2$, and the completely
 synchronous solution is stable for $r<(\cos\alpha)/2$.
 Black circles denote the parameter values of Fig.~\ref{fig: chimera
 states}.
 Black triangles denote the parameter values of Fig.~\ref{fig:
 multichimera states}, where multichimera states appear.
 }
\end{figure}
Investigating the chimera states in the $(\cos\alpha,r)$ parameter
space, we obtained Fig.~\ref{fig: stability region}.
The red region corresponding to the chimera state in the continuous
limit ($N=2000$) is spread around $r=(\cos\alpha)/2$.
In the finite $N$ cases, the chimera state for small $r$ becomes
transient, while the chimera state for large $r$ remains persistent, as
seen at least for $r>(\cos\alpha)/2$ of the red region.
For $\cos\alpha<0.15$, we can see that there exists a region
$r^{\ast}<r<(\cos\alpha)/2$ where the chimera state is persistent in the
case of $N=30$.
Note that the stability region of the persistent chimera state (hatched
in Fig.~\ref{fig: stability region}) extends to the $r=0$ line, which
implies that the chimera state with the sine coupling can be persistent
(non-transient) even in the finite $N$ cases.
Specifically, the average lifetime $\tau$ of the chimera state increases
similarly to Fig.~\ref{fig: average lifetime} as $\cos\alpha$ is
decreased on the $r=0$ line, and diverge at
$\cos\alpha^{\ast}\simeq0.044$.
However, this fact does not contradict the previous study that shows the
transient chimera state~\cite{WO2011}, because the parameter $\alpha$ in
that study corresponds to the line of black circles in Fig.~\ref{fig:
stability region}, which has a larger $\cos\alpha$ than our hatched
region on the $r=0$ line.

\begin{figure}[!b]
 \centering
 \includegraphics[width=86truemm]{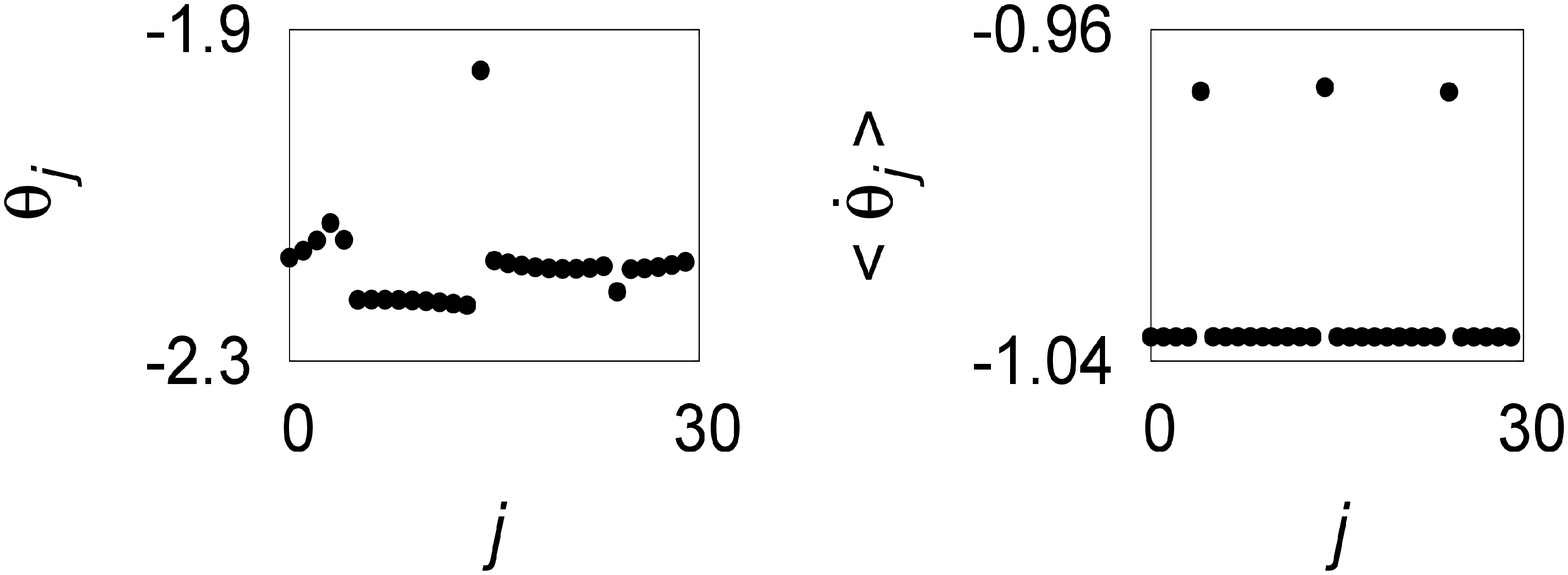}
 \caption{\label{fig: weak chimera}
 Weak chimera for Eq.~(\ref{eq: phase oscillators}) with $N=30$,
 $\alpha=1.46$, and $r=0.032$, which are the parameter values on the
 line of black circles in Fig.~\ref{fig: stability region}.
 The snapshot of phase $\theta_j$ (left) and the profile of the average
 frequency $\langle\dot{\theta}_j\rangle$ (right).
 }
 \vspace{5truemm}
 \centering
 \includegraphics[width=86truemm]{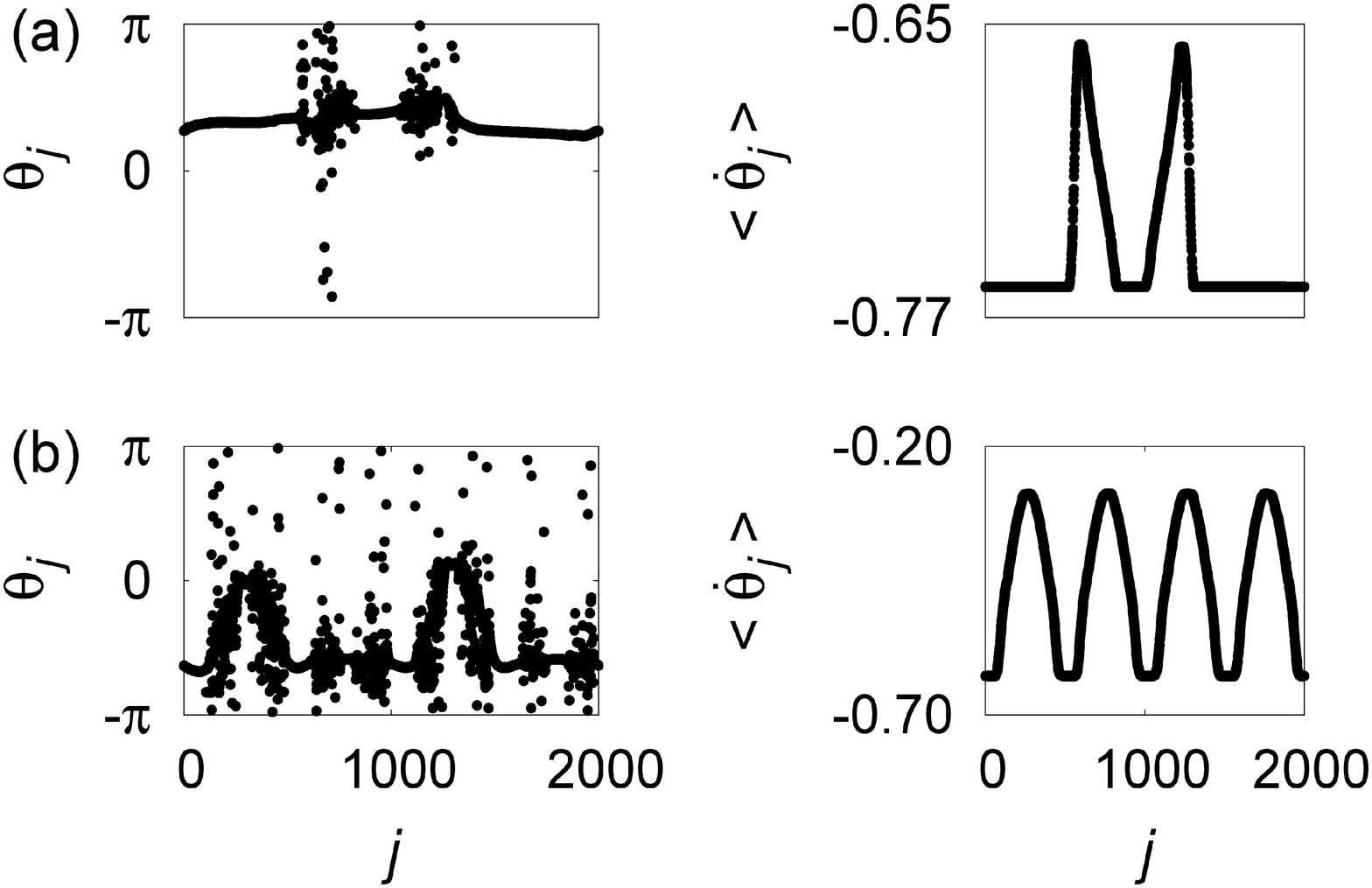}
 \caption{\label{fig: multichimera states}
 Multichimera states for Eq.~(\ref{eq: phase oscillators}) with
 $N=2000$.
 The left panels show the snapshot of phase $\theta_j$, and the right
 panels show the profile of the average frequency
 $\langle\dot{\theta}_j\rangle$ with $T=5000$ and $t_{\text{rel}}=2000$
 in Eq.~(\ref{eq: average frequency}).
 Parameter values are (a)~$\alpha=0.95$ and $r=0.28$, and
 (b)~$\alpha=1.22$ and $r=0.30$, which are plotted in Fig.~\ref{fig:
 stability region}.
 }
\end{figure}
In summary, we studied chimera states in the systems of nonlocally
coupled phase oscillators, Eq.~(\ref{eq: phase oscillators}), with the
Hansel-Mato-Meunier coupling, Eq.~(\ref{eq: HMM coupling}), by numerical
simulations, motivated by the result that chimera states with the sine
coupling, Eq.~(\ref{eq: sine coupling}), in finitely discretized systems
are chaotic transient and finally collapse into the completely
synchronous state~\cite{WO2011}.
The existence of chimera states was examined in the parameter space
($\alpha,r$) in Eq.~(\ref{eq: HMM coupling}), and the chimera states
were observed around $r=(\cos\alpha)/2$ in the continuous limit
$N\rightarrow\infty$.
For $r<(\cos\alpha)/2$, the chimera state and the completely synchronous
state can be bistable.
In this region of the finite $N$ cases, the chimera state is transient
for $r<r^{\ast}$, but it is persistent for $r^{\ast}<r<(\cos\alpha)/2$.
Moreover, even for $r>(\cos\alpha)/2$, it is persistent in the region
where the chimera state in $N\rightarrow\infty$ is stable.
At first, we expected the chimera state to become persistent due to the
destabilization of the completely synchronous state by the effect of
$r$, but have obtained the persistent chimera state not only in the
unstable region of the completely synchronous state as expected but also
in its stable region.
As a result, we have discovered that the chimera state in the case of
the sine coupling can also be persistent by using appropriate $\alpha$
in the stability region of the completely synchronous state.
Though we have numerically found the persistent chimera state in this
Rapid Communication, its bifurcation-theoretical understanding is still
an open problem.

When we investigated the collapse of chimera states at $\alpha=1.46$, we
infrequently observed that a chimera state collapses into a weak chimera
characterized by the coexistence of frequency-synchronous and
-asynchronous oscillators~\cite{AB2015, BA2015}, as shown in
Fig.~\ref{fig: weak chimera}.
In~\cite{AB2015}, the existence of weak chimeras for Eq.~(\ref{eq: phase
oscillators}) with Eq.~(\ref{eq: HMM coupling}) is confirmed in the
system with a small number of oscillators ($N=4$, $6$, and $10$).
In our numerical simulation with a larger number of oscillators
($N=30$), such a weak chimera is stable in a small range of
$r<(\cos\alpha)/2$, for example, $0.032\leq r\leq0.040$ at
$\alpha=1.46$.

Moreover, as other solutions, we observed multichimera states, which
have two or more incoherent domains~\cite{OOHS2013, SSA2008, MVSLM2014,
XKK2014}, for Eq.~(\ref{eq: phase oscillators}) with Eq.~(\ref{eq: HMM
coupling}) in the continuous limit ($N=2000$), as shown in
Fig.~\ref{fig: multichimera states}.
Other than the black triangles in Fig.~\ref{fig: stability region}, we
observed multichimera states in a large region of the parameter space,
though we do not describe the region in detail because it is beyond the
scope of the present Rapid Communication.

\bibliography{reference}

\begin{thebibliography}{35}%
\makeatletter
\providecommand \@ifxundefined [1]{%
 \@ifx{#1\undefined}
}%
\providecommand \@ifnum [1]{%
 \ifnum #1\expandafter \@firstoftwo
 \else \expandafter \@secondoftwo
 \fi
}%
\providecommand \@ifx [1]{%
 \ifx #1\expandafter \@firstoftwo
 \else \expandafter \@secondoftwo
 \fi
}%
\providecommand \natexlab [1]{#1}%
\providecommand \enquote  [1]{``#1''}%
\providecommand \bibnamefont  [1]{#1}%
\providecommand \bibfnamefont [1]{#1}%
\providecommand \citenamefont [1]{#1}%
\providecommand \href@noop [0]{\@secondoftwo}%
\providecommand \href [0]{\begingroup \@sanitize@url \@href}%
\providecommand \@href[1]{\@@startlink{#1}\@@href}%
\providecommand \@@href[1]{\endgroup#1\@@endlink}%
\providecommand \@sanitize@url [0]{\catcode `\\12\catcode `\$12\catcode
  `\&12\catcode `\#12\catcode `\^12\catcode `\_12\catcode `\%12\relax}%
\providecommand \@@startlink[1]{}%
\providecommand \@@endlink[0]{}%
\providecommand \url  [0]{\begingroup\@sanitize@url \@url }%
\providecommand \@url [1]{\endgroup\@href {#1}{\urlprefix }}%
\providecommand \urlprefix  [0]{URL }%
\providecommand \Eprint [0]{\href }%
\providecommand \doibase [0]{http://dx.doi.org/}%
\providecommand \selectlanguage [0]{\@gobble}%
\providecommand \bibinfo  [0]{\@secondoftwo}%
\providecommand \bibfield  [0]{\@secondoftwo}%
\providecommand \translation [1]{[#1]}%
\providecommand \BibitemOpen [0]{}%
\providecommand \bibitemStop [0]{}%
\providecommand \bibitemNoStop [0]{.\EOS\space}%
\providecommand \EOS [0]{\spacefactor3000\relax}%
\providecommand \BibitemShut  [1]{\csname bibitem#1\endcsname}%
\let\auto@bib@innerbib\@empty
\bibitem [{\citenamefont {Kuramoto}(1984)}]{b:Kuramoto}%
  \BibitemOpen
  \bibfield  {author} {\bibinfo {author} {\bibfnamefont {Y.}~\bibnamefont
  {Kuramoto}},\ }\href@noop {} {\emph {\bibinfo {title} {Chemical Oscillation,
  Waves, and Turbulence}}}\ (\bibinfo  {publisher} {Springer},\ \bibinfo
  {address} {Berlin},\ \bibinfo {year} {1984})\BibitemShut {NoStop}%
\bibitem [{\citenamefont {Pikovsky}\ \emph {et~al.}(2003)\citenamefont
  {Pikovsky}, \citenamefont {Rosenblum},\ and\ \citenamefont {Kurths}}]{b:PRK}%
  \BibitemOpen
  \bibfield  {author} {\bibinfo {author} {\bibfnamefont {A.}~\bibnamefont
  {Pikovsky}}, \bibinfo {author} {\bibfnamefont {M.}~\bibnamefont {Rosenblum}},
  \ and\ \bibinfo {author} {\bibfnamefont {J.}~\bibnamefont {Kurths}},\
  }\href@noop {} {\emph {\bibinfo {title} {Synchronization: A Universal Concept
  in Nonlinear Sciences}}}\ (\bibinfo  {publisher} {Cambridge University
  Press},\ \bibinfo {address} {Cambridge},\ \bibinfo {year} {2003})\BibitemShut
  {NoStop}%
\bibitem [{\citenamefont {Kuramoto}\ and\ \citenamefont
  {Battogtokh}(2002)}]{KB2002}%
  \BibitemOpen
  \bibfield  {author} {\bibinfo {author} {\bibfnamefont {Y.}~\bibnamefont
  {Kuramoto}}\ and\ \bibinfo {author} {\bibfnamefont {D.}~\bibnamefont
  {Battogtokh}},\ }\href@noop {} {\bibfield  {journal} {\bibinfo  {journal}
  {Nonlinear Phenom. Complex Syst.}\ }\textbf {\bibinfo {volume} {5}},\
  \bibinfo {pages} {380} (\bibinfo {year} {2002})}\BibitemShut {NoStop}%
\bibitem [{\citenamefont {Abrams}\ and\ \citenamefont
  {Strogatz}(2004)}]{AS2004}%
  \BibitemOpen
  \bibfield  {author} {\bibinfo {author} {\bibfnamefont {D.~M.}\ \bibnamefont
  {Abrams}}\ and\ \bibinfo {author} {\bibfnamefont {S.~H.}\ \bibnamefont
  {Strogatz}},\ }\href@noop {} {\bibfield  {journal} {\bibinfo  {journal}
  {Phys. Rev. Lett.}\ }\textbf {\bibinfo {volume} {93}},\ \bibinfo {pages}
  {174102} (\bibinfo {year} {2004})}\BibitemShut {NoStop}%
\bibitem [{\citenamefont {Shima}\ and\ \citenamefont
  {Kuramoto}(2004)}]{SK2004}%
  \BibitemOpen
  \bibfield  {author} {\bibinfo {author} {\bibfnamefont {S.~I.}\ \bibnamefont
  {Shima}}\ and\ \bibinfo {author} {\bibfnamefont {Y.}~\bibnamefont
  {Kuramoto}},\ }\href@noop {} {\bibfield  {journal} {\bibinfo  {journal}
  {Phys. Rev. E}\ }\textbf {\bibinfo {volume} {69}},\ \bibinfo {pages} {036213}
  (\bibinfo {year} {2004})}\BibitemShut {NoStop}%
\bibitem [{\citenamefont {Abrams}\ and\ \citenamefont
  {Strogatz}(2006)}]{AS2006}%
  \BibitemOpen
  \bibfield  {author} {\bibinfo {author} {\bibfnamefont {D.~M.}\ \bibnamefont
  {Abrams}}\ and\ \bibinfo {author} {\bibfnamefont {S.~H.}\ \bibnamefont
  {Strogatz}},\ }\href@noop {} {\bibfield  {journal} {\bibinfo  {journal} {Int.
  J. Bifurcation Chaos}\ }\textbf {\bibinfo {volume} {16}},\ \bibinfo {pages}
  {21} (\bibinfo {year} {2006})}\BibitemShut {NoStop}%
\bibitem [{\citenamefont {Abrams}\ \emph {et~al.}(2008)\citenamefont {Abrams},
  \citenamefont {Mirollo}, \citenamefont {Strogatz},\ and\ \citenamefont
  {Wiley}}]{AMSW2008}%
  \BibitemOpen
  \bibfield  {author} {\bibinfo {author} {\bibfnamefont {D.~M.}\ \bibnamefont
  {Abrams}}, \bibinfo {author} {\bibfnamefont {R.}~\bibnamefont {Mirollo}},
  \bibinfo {author} {\bibfnamefont {S.~H.}\ \bibnamefont {Strogatz}}, \ and\
  \bibinfo {author} {\bibfnamefont {D.~A.}\ \bibnamefont {Wiley}},\ }\href@noop
  {} {\bibfield  {journal} {\bibinfo  {journal} {Phys. Rev. Lett.}\ }\textbf
  {\bibinfo {volume} {101}},\ \bibinfo {pages} {084103} (\bibinfo {year}
  {2008})}\BibitemShut {NoStop}%
\bibitem [{\citenamefont {Ott}\ and\ \citenamefont {Antonsen}(2008)}]{OA2008}%
  \BibitemOpen
  \bibfield  {author} {\bibinfo {author} {\bibfnamefont {E.}~\bibnamefont
  {Ott}}\ and\ \bibinfo {author} {\bibfnamefont {T.~M.}\ \bibnamefont
  {Antonsen}},\ }\href@noop {} {\bibfield  {journal} {\bibinfo  {journal}
  {Chaos}\ }\textbf {\bibinfo {volume} {18}},\ \bibinfo {pages} {037113}
  (\bibinfo {year} {2008})}\BibitemShut {NoStop}%
\bibitem [{\citenamefont {Ott}\ and\ \citenamefont {Antonsen}(2009)}]{OA2009}%
  \BibitemOpen
  \bibfield  {author} {\bibinfo {author} {\bibfnamefont {E.}~\bibnamefont
  {Ott}}\ and\ \bibinfo {author} {\bibfnamefont {T.~M.}\ \bibnamefont
  {Antonsen}},\ }\href@noop {} {\bibfield  {journal} {\bibinfo  {journal}
  {Chaos}\ }\textbf {\bibinfo {volume} {19}},\ \bibinfo {pages} {023117}
  (\bibinfo {year} {2009})}\BibitemShut {NoStop}%
\bibitem [{\citenamefont {Martens}\ \emph {et~al.}(2010)\citenamefont
  {Martens}, \citenamefont {Laing},\ and\ \citenamefont {Strogatz}}]{MLS2010}%
  \BibitemOpen
  \bibfield  {author} {\bibinfo {author} {\bibfnamefont {E.~A.}\ \bibnamefont
  {Martens}}, \bibinfo {author} {\bibfnamefont {C.~R.}\ \bibnamefont {Laing}},
  \ and\ \bibinfo {author} {\bibfnamefont {S.~H.}\ \bibnamefont {Strogatz}},\
  }\href@noop {} {\bibfield  {journal} {\bibinfo  {journal} {Phys. Rev. Lett.}\
  }\textbf {\bibinfo {volume} {104}},\ \bibinfo {pages} {044101} (\bibinfo
  {year} {2010})}\BibitemShut {NoStop}%
\bibitem [{\citenamefont {Omel'chenko}\ \emph {et~al.}(2010)\citenamefont
  {Omel'chenko}, \citenamefont {Wolfrum},\ and\ \citenamefont
  {Maistrenko}}]{OWM2010}%
  \BibitemOpen
  \bibfield  {author} {\bibinfo {author} {\bibfnamefont {O.~E.}\ \bibnamefont
  {Omel'chenko}}, \bibinfo {author} {\bibfnamefont {M.}~\bibnamefont
  {Wolfrum}}, \ and\ \bibinfo {author} {\bibfnamefont {Y.~L.}\ \bibnamefont
  {Maistrenko}},\ }\href@noop {} {\bibfield  {journal} {\bibinfo  {journal}
  {Phys. Rev. E}\ }\textbf {\bibinfo {volume} {81}},\ \bibinfo {pages}
  {065201(R)} (\bibinfo {year} {2010})}\BibitemShut {NoStop}%
\bibitem [{\citenamefont {Wolfrum}\ \emph {et~al.}(2011)\citenamefont
  {Wolfrum}, \citenamefont {Omel'chenko}, \citenamefont {Yanchuk},\ and\
  \citenamefont {Maistrenko}}]{WOYM2011}%
  \BibitemOpen
  \bibfield  {author} {\bibinfo {author} {\bibfnamefont {M.}~\bibnamefont
  {Wolfrum}}, \bibinfo {author} {\bibfnamefont {O.~E.}\ \bibnamefont
  {Omel'chenko}}, \bibinfo {author} {\bibfnamefont {S.}~\bibnamefont
  {Yanchuk}}, \ and\ \bibinfo {author} {\bibfnamefont {Y.~L.}\ \bibnamefont
  {Maistrenko}},\ }\href@noop {} {\bibfield  {journal} {\bibinfo  {journal}
  {Chaos}\ }\textbf {\bibinfo {volume} {21}},\ \bibinfo {pages} {013112}
  (\bibinfo {year} {2011})}\BibitemShut {NoStop}%
\bibitem [{\citenamefont {Wolfrum}\ and\ \citenamefont
  {Omel'chenko}(2011)}]{WO2011}%
  \BibitemOpen
  \bibfield  {author} {\bibinfo {author} {\bibfnamefont {M.}~\bibnamefont
  {Wolfrum}}\ and\ \bibinfo {author} {\bibfnamefont {O.~E.}\ \bibnamefont
  {Omel'chenko}},\ }\href@noop {} {\bibfield  {journal} {\bibinfo  {journal}
  {Phys. Rev. E}\ }\textbf {\bibinfo {volume} {84}},\ \bibinfo {pages}
  {015201(R)} (\bibinfo {year} {2011})}\BibitemShut {NoStop}%
\bibitem [{\citenamefont {Omelchenko}\ \emph {et~al.}(2011)\citenamefont
  {Omelchenko}, \citenamefont {Maistrenko}, \citenamefont {H{\"o}vel},\ and\
  \citenamefont {Sch{\"o}ll}}]{OMHS2011}%
  \BibitemOpen
  \bibfield  {author} {\bibinfo {author} {\bibfnamefont {I.}~\bibnamefont
  {Omelchenko}}, \bibinfo {author} {\bibfnamefont {Y.}~\bibnamefont
  {Maistrenko}}, \bibinfo {author} {\bibfnamefont {P.}~\bibnamefont
  {H{\"o}vel}}, \ and\ \bibinfo {author} {\bibfnamefont {E.}~\bibnamefont
  {Sch{\"o}ll}},\ }\href@noop {} {\bibfield  {journal} {\bibinfo  {journal}
  {Phys. Rev. Lett.}\ }\textbf {\bibinfo {volume} {106}},\ \bibinfo {pages}
  {234102} (\bibinfo {year} {2011})}\BibitemShut {NoStop}%
\bibitem [{\citenamefont {Omelchenko}\ \emph {et~al.}(2012)\citenamefont
  {Omelchenko}, \citenamefont {Riemenschneider}, \citenamefont {H{\"o}vel},
  \citenamefont {Maistrenko},\ and\ \citenamefont {Sch{\"o}ll}}]{ORHMS2012}%
  \BibitemOpen
  \bibfield  {author} {\bibinfo {author} {\bibfnamefont {I.}~\bibnamefont
  {Omelchenko}}, \bibinfo {author} {\bibfnamefont {B.}~\bibnamefont
  {Riemenschneider}}, \bibinfo {author} {\bibfnamefont {P.}~\bibnamefont
  {H{\"o}vel}}, \bibinfo {author} {\bibfnamefont {Y.}~\bibnamefont
  {Maistrenko}}, \ and\ \bibinfo {author} {\bibfnamefont {E.}~\bibnamefont
  {Sch{\"o}ll}},\ }\href@noop {} {\bibfield  {journal} {\bibinfo  {journal}
  {Phys. Rev. E}\ }\textbf {\bibinfo {volume} {85}},\ \bibinfo {pages} {026212}
  (\bibinfo {year} {2012})}\BibitemShut {NoStop}%
\bibitem [{\citenamefont {Tinsley}\ \emph {et~al.}(2012)\citenamefont
  {Tinsley}, \citenamefont {Nkomo},\ and\ \citenamefont {Showalter}}]{TNS2012}%
  \BibitemOpen
  \bibfield  {author} {\bibinfo {author} {\bibfnamefont {M.~R.}\ \bibnamefont
  {Tinsley}}, \bibinfo {author} {\bibfnamefont {S.}~\bibnamefont {Nkomo}}, \
  and\ \bibinfo {author} {\bibfnamefont {K.}~\bibnamefont {Showalter}},\
  }\href@noop {} {\bibfield  {journal} {\bibinfo  {journal} {Nat. Phys.}\
  }\textbf {\bibinfo {volume} {8}},\ \bibinfo {pages} {662} (\bibinfo {year}
  {2012})}\BibitemShut {NoStop}%
\bibitem [{\citenamefont {Hagerstrom}\ \emph {et~al.}(2012)\citenamefont
  {Hagerstrom}, \citenamefont {Murphy}, \citenamefont {Roy}, \citenamefont
  {H{\"o}vel}, \citenamefont {Omelchenko},\ and\ \citenamefont
  {Sch{\"o}ll}}]{HMRHOS2012}%
  \BibitemOpen
  \bibfield  {author} {\bibinfo {author} {\bibfnamefont {A.~M.}\ \bibnamefont
  {Hagerstrom}}, \bibinfo {author} {\bibfnamefont {T.~E.}\ \bibnamefont
  {Murphy}}, \bibinfo {author} {\bibfnamefont {R.}~\bibnamefont {Roy}},
  \bibinfo {author} {\bibfnamefont {P.}~\bibnamefont {H{\"o}vel}}, \bibinfo
  {author} {\bibfnamefont {I.}~\bibnamefont {Omelchenko}}, \ and\ \bibinfo
  {author} {\bibfnamefont {E.}~\bibnamefont {Sch{\"o}ll}},\ }\href@noop {}
  {\bibfield  {journal} {\bibinfo  {journal} {Nature Phys.}\ }\textbf {\bibinfo
  {volume} {8}},\ \bibinfo {pages} {658} (\bibinfo {year} {2012})}\BibitemShut
  {NoStop}%
\bibitem [{\citenamefont {Omelchenko}\ \emph {et~al.}(2013)\citenamefont
  {Omelchenko}, \citenamefont {Omel'chenko}, \citenamefont {H{\"o}vel},\ and\
  \citenamefont {Sch{\"o}ll}}]{OOHS2013}%
  \BibitemOpen
  \bibfield  {author} {\bibinfo {author} {\bibfnamefont {I.}~\bibnamefont
  {Omelchenko}}, \bibinfo {author} {\bibfnamefont {O.~E.}\ \bibnamefont
  {Omel'chenko}}, \bibinfo {author} {\bibfnamefont {P.}~\bibnamefont
  {H{\"o}vel}}, \ and\ \bibinfo {author} {\bibfnamefont {E.}~\bibnamefont
  {Sch{\"o}ll}},\ }\href@noop {} {\bibfield  {journal} {\bibinfo  {journal}
  {Phys. Rev. Lett.}\ }\textbf {\bibinfo {volume} {110}},\ \bibinfo {pages}
  {224101} (\bibinfo {year} {2013})}\BibitemShut {NoStop}%
\bibitem [{\citenamefont {Martens}\ \emph {et~al.}(2013)\citenamefont
  {Martens}, \citenamefont {Thutupalli}, \citenamefont {Fourri{\'e}re},\ and\
  \citenamefont {Hallatschek}}]{MTFH2013}%
  \BibitemOpen
  \bibfield  {author} {\bibinfo {author} {\bibfnamefont {E.~A.}\ \bibnamefont
  {Martens}}, \bibinfo {author} {\bibfnamefont {S.}~\bibnamefont {Thutupalli}},
  \bibinfo {author} {\bibfnamefont {A.}~\bibnamefont {Fourri{\'e}re}}, \ and\
  \bibinfo {author} {\bibfnamefont {O.}~\bibnamefont {Hallatschek}},\
  }\href@noop {} {\bibfield  {journal} {\bibinfo  {journal} {Proc. Natl. Acad.
  Sci. U.S.A.}\ }\textbf {\bibinfo {volume} {110}},\ \bibinfo {pages} {10563}
  (\bibinfo {year} {2013})}\BibitemShut {NoStop}%
\bibitem [{\citenamefont {Schmidt}\ \emph {et~al.}(2014)\citenamefont
  {Schmidt}, \citenamefont {Sch{\"o}nleber}, \citenamefont {Krischer},\ and\
  \citenamefont {Garc{\'i}a-Morales}}]{SSKG2014}%
  \BibitemOpen
  \bibfield  {author} {\bibinfo {author} {\bibfnamefont {L.}~\bibnamefont
  {Schmidt}}, \bibinfo {author} {\bibfnamefont {K.}~\bibnamefont
  {Sch{\"o}nleber}}, \bibinfo {author} {\bibfnamefont {K.}~\bibnamefont
  {Krischer}}, \ and\ \bibinfo {author} {\bibfnamefont {V.}~\bibnamefont
  {Garc{\'i}a-Morales}},\ }\href@noop {} {\bibfield  {journal} {\bibinfo
  {journal} {Chaos}\ }\textbf {\bibinfo {volume} {24}},\ \bibinfo {pages}
  {013102} (\bibinfo {year} {2014})}\BibitemShut {NoStop}%
\bibitem [{\citenamefont {Zhu}\ \emph {et~al.}(2014)\citenamefont {Zhu},
  \citenamefont {Zheng},\ and\ \citenamefont {Yang}}]{ZZY2014}%
  \BibitemOpen
  \bibfield  {author} {\bibinfo {author} {\bibfnamefont {Y.}~\bibnamefont
  {Zhu}}, \bibinfo {author} {\bibfnamefont {Z.}~\bibnamefont {Zheng}}, \ and\
  \bibinfo {author} {\bibfnamefont {J.}~\bibnamefont {Yang}},\ }\href@noop {}
  {\bibfield  {journal} {\bibinfo  {journal} {Phys. Rev. E}\ }\textbf {\bibinfo
  {volume} {89}},\ \bibinfo {pages} {022914} (\bibinfo {year}
  {2014})}\BibitemShut {NoStop}%
\bibitem [{\citenamefont {Haugland}\ \emph {et~al.}(2015)\citenamefont
  {Haugland}, \citenamefont {Schmidt},\ and\ \citenamefont
  {Krischer}}]{HSK2015}%
  \BibitemOpen
  \bibfield  {author} {\bibinfo {author} {\bibfnamefont {S.~W.}\ \bibnamefont
  {Haugland}}, \bibinfo {author} {\bibfnamefont {L.}~\bibnamefont {Schmidt}}, \
  and\ \bibinfo {author} {\bibfnamefont {K.}~\bibnamefont {Krischer}},\
  }\href@noop {} {\bibfield  {journal} {\bibinfo  {journal} {Sci. Rep.}\
  }\textbf {\bibinfo {volume} {5}},\ \bibinfo {pages} {9883} (\bibinfo {year}
  {2015})}\BibitemShut {NoStop}%
\bibitem [{\citenamefont {Sethia}\ \emph {et~al.}(2008)\citenamefont {Sethia},
  \citenamefont {Sen},\ and\ \citenamefont {Atay}}]{SSA2008}%
  \BibitemOpen
  \bibfield  {author} {\bibinfo {author} {\bibfnamefont {G.~C.}\ \bibnamefont
  {Sethia}}, \bibinfo {author} {\bibfnamefont {A.}~\bibnamefont {Sen}}, \ and\
  \bibinfo {author} {\bibfnamefont {F.~M.}\ \bibnamefont {Atay}},\ }\href@noop
  {} {\bibfield  {journal} {\bibinfo  {journal} {Phys. Rev. Lett.}\ }\textbf
  {\bibinfo {volume} {100}},\ \bibinfo {pages} {144102} (\bibinfo {year}
  {2008})}\BibitemShut {NoStop}%
\bibitem [{\citenamefont {Maistrenko}\ \emph {et~al.}(2014)\citenamefont
  {Maistrenko}, \citenamefont {Vasylenko}, \citenamefont {Sudakov},
  \citenamefont {Levchenko},\ and\ \citenamefont {Maistrenko}}]{MVSLM2014}%
  \BibitemOpen
  \bibfield  {author} {\bibinfo {author} {\bibfnamefont {Y.~L.}\ \bibnamefont
  {Maistrenko}}, \bibinfo {author} {\bibfnamefont {A.}~\bibnamefont
  {Vasylenko}}, \bibinfo {author} {\bibfnamefont {O.}~\bibnamefont {Sudakov}},
  \bibinfo {author} {\bibfnamefont {R.}~\bibnamefont {Levchenko}}, \ and\
  \bibinfo {author} {\bibfnamefont {V.~L.}\ \bibnamefont {Maistrenko}},\
  }\href@noop {} {\bibfield  {journal} {\bibinfo  {journal} {Int. J.
  Bifurcation Chaos}\ }\textbf {\bibinfo {volume} {24}},\ \bibinfo {pages}
  {1440014} (\bibinfo {year} {2014})}\BibitemShut {NoStop}%
\bibitem [{\citenamefont {Xie}\ \emph {et~al.}(2014)\citenamefont {Xie},
  \citenamefont {Knobloch},\ and\ \citenamefont {Kao}}]{XKK2014}%
  \BibitemOpen
  \bibfield  {author} {\bibinfo {author} {\bibfnamefont {J.}~\bibnamefont
  {Xie}}, \bibinfo {author} {\bibfnamefont {E.}~\bibnamefont {Knobloch}}, \
  and\ \bibinfo {author} {\bibfnamefont {H.-C.}\ \bibnamefont {Kao}},\
  }\href@noop {} {\bibfield  {journal} {\bibinfo  {journal} {Phys. Rev. E}\
  }\textbf {\bibinfo {volume} {90}},\ \bibinfo {pages} {022919} (\bibinfo
  {year} {2014})}\BibitemShut {NoStop}%
\bibitem [{\citenamefont {Rosin}\ \emph {et~al.}(2014)\citenamefont {Rosin},
  \citenamefont {Rontani}, \citenamefont {Haynes}, \citenamefont {Sch{\"o}ll},\
  and\ \citenamefont {Gauthier}}]{RRHSG2014}%
  \BibitemOpen
  \bibfield  {author} {\bibinfo {author} {\bibfnamefont {D.~P.}\ \bibnamefont
  {Rosin}}, \bibinfo {author} {\bibfnamefont {D.}~\bibnamefont {Rontani}},
  \bibinfo {author} {\bibfnamefont {N.~D.}\ \bibnamefont {Haynes}}, \bibinfo
  {author} {\bibfnamefont {E.}~\bibnamefont {Sch{\"o}ll}}, \ and\ \bibinfo
  {author} {\bibfnamefont {D.~J.}\ \bibnamefont {Gauthier}},\ }\href@noop {}
  {\bibfield  {journal} {\bibinfo  {journal} {Phys. Rev. E}\ }\textbf {\bibinfo
  {volume} {90}},\ \bibinfo {pages} {030902(R)} (\bibinfo {year}
  {2014})}\BibitemShut {NoStop}%
\bibitem [{\citenamefont {Sakaguchi}\ and\ \citenamefont
  {Kuramoto}(1986)}]{SK1986}%
  \BibitemOpen
  \bibfield  {author} {\bibinfo {author} {\bibfnamefont {H.}~\bibnamefont
  {Sakaguchi}}\ and\ \bibinfo {author} {\bibfnamefont {Y.}~\bibnamefont
  {Kuramoto}},\ }\href@noop {} {\bibfield  {journal} {\bibinfo  {journal}
  {Prog. Theor. Phys.}\ }\textbf {\bibinfo {volume} {76}},\ \bibinfo {pages}
  {576} (\bibinfo {year} {1986})}\BibitemShut {NoStop}%
\bibitem [{\citenamefont {Ashwin}\ and\ \citenamefont
  {Burylko}(2015)}]{AB2015}%
  \BibitemOpen
  \bibfield  {author} {\bibinfo {author} {\bibfnamefont {P.}~\bibnamefont
  {Ashwin}}\ and\ \bibinfo {author} {\bibfnamefont {O.}~\bibnamefont
  {Burylko}},\ }\href@noop {} {\bibfield  {journal} {\bibinfo  {journal}
  {Chaos}\ }\textbf {\bibinfo {volume} {25}},\ \bibinfo {pages} {013106}
  (\bibinfo {year} {2015})}\BibitemShut {NoStop}%
\bibitem [{\citenamefont {Hansel}\ \emph {et~al.}(1993)\citenamefont {Hansel},
  \citenamefont {Mato},\ and\ \citenamefont {Meunier}}]{HMM1993}%
  \BibitemOpen
  \bibfield  {author} {\bibinfo {author} {\bibfnamefont {D.}~\bibnamefont
  {Hansel}}, \bibinfo {author} {\bibfnamefont {G.}~\bibnamefont {Mato}}, \ and\
  \bibinfo {author} {\bibfnamefont {C.}~\bibnamefont {Meunier}},\ }\href@noop
  {} {\bibfield  {journal} {\bibinfo  {journal} {Phys. Rev. E}\ }\textbf
  {\bibinfo {volume} {48}},\ \bibinfo {pages} {3470} (\bibinfo {year}
  {1993})}\BibitemShut {NoStop}%
\bibitem [{\citenamefont {Daido}(1992)}]{Daido1992}%
  \BibitemOpen
  \bibfield  {author} {\bibinfo {author} {\bibfnamefont {H.}~\bibnamefont
  {Daido}},\ }\href@noop {} {\bibfield  {journal} {\bibinfo  {journal} {Prog.
  Theor. Phys.}\ }\textbf {\bibinfo {volume} {88}},\ \bibinfo {pages} {1213}
  (\bibinfo {year} {1992})}\BibitemShut {NoStop}%
\bibitem [{\citenamefont {Okuda}(1993)}]{Okuda1993}%
  \BibitemOpen
  \bibfield  {author} {\bibinfo {author} {\bibfnamefont {K.}~\bibnamefont
  {Okuda}},\ }\href@noop {} {\bibfield  {journal} {\bibinfo  {journal} {Physica
  D}\ }\textbf {\bibinfo {volume} {63}},\ \bibinfo {pages} {424} (\bibinfo
  {year} {1993})}\BibitemShut {NoStop}%
\bibitem [{\citenamefont {Daido}(1996)}]{Daido1996}%
  \BibitemOpen
  \bibfield  {author} {\bibinfo {author} {\bibfnamefont {H.}~\bibnamefont
  {Daido}},\ }\href@noop {} {\bibfield  {journal} {\bibinfo  {journal} {Physica
  D}\ }\textbf {\bibinfo {volume} {91}},\ \bibinfo {pages} {24} (\bibinfo
  {year} {1996})}\BibitemShut {NoStop}%
\bibitem [{\citenamefont {Kori}\ and\ \citenamefont {Kuramoto}(2001)}]{KK2001}%
  \BibitemOpen
  \bibfield  {author} {\bibinfo {author} {\bibfnamefont {H.}~\bibnamefont
  {Kori}}\ and\ \bibinfo {author} {\bibfnamefont {Y.}~\bibnamefont
  {Kuramoto}},\ }\href@noop {} {\bibfield  {journal} {\bibinfo  {journal}
  {Phys. Rev. E}\ }\textbf {\bibinfo {volume} {63}},\ \bibinfo {pages} {046214}
  (\bibinfo {year} {2001})}\BibitemShut {NoStop}%
\bibitem [{\citenamefont {Ashwin}\ \emph {et~al.}(2008)\citenamefont {Ashwin},
  \citenamefont {Burylko},\ and\ \citenamefont {Maistrenko}}]{ABM2008}%
  \BibitemOpen
  \bibfield  {author} {\bibinfo {author} {\bibfnamefont {P.}~\bibnamefont
  {Ashwin}}, \bibinfo {author} {\bibfnamefont {O.}~\bibnamefont {Burylko}}, \
  and\ \bibinfo {author} {\bibfnamefont {Y.}~\bibnamefont {Maistrenko}},\
  }\href@noop {} {\bibfield  {journal} {\bibinfo  {journal} {Physica D}\
  }\textbf {\bibinfo {volume} {237}},\ \bibinfo {pages} {454–466} (\bibinfo
  {year} {2008})}\BibitemShut {NoStop}%
\bibitem [{\citenamefont {Bick}\ and\ \citenamefont {Ashwin}(2015)}]{BA2015}%
  \BibitemOpen
  \bibfield  {author} {\bibinfo {author} {\bibfnamefont {C.}~\bibnamefont
  {Bick}}\ and\ \bibinfo {author} {\bibfnamefont {P.}~\bibnamefont {Ashwin}},\
  }\href@noop {} {\bibfield  {journal} {\bibinfo  {journal} {arXiv:1509.08824}\
  } (\bibinfo {year} {2015})}\BibitemShut {NoStop}%
\end{thebibliography}%

\end{document}